\documentclass[prl,aps,twocolumn,showpacs,groupedaddress,superscriptaddress,amsmath,amssymb,10pt]{revtex4}
\usepackage{url,graphicx, color}
\usepackage{dcolumn}
\usepackage{bm}
\usepackage{epsfig,amsmath}
\usepackage{amssymb,bm}
\usepackage{color}

\def\pp{p}

\begin{document}

\title{Criticality in Dynamic Arrest: Correspondence between Glasses and Traffic}

\author{A.~S.~de Wijn}
\email{astrid@dewijn.eu}
\author{D.~M.~Miedema}
\email{D.M.Miedema@uva.nl}
\author{B.~Nienhuis}
\email{B.Nienhuis@uva.nl}
\author{P.~Schall}
\email{ps@peterschall.de}
\affiliation{Institute of Physics, University of Amsterdam, P.O.~box 94485, 1090 GL Amsterdam, The Netherlands}

\begin{abstract}
Dynamic arrest is a general phenomenon across a wide range of dynamic systems, but the universality of dynamic arrest phenomena remains unclear. We relate the emergence of traffic jams in a simple traffic flow model to the dynamic slow down in kinetically constrained models for glasses. In kinetically constrained models, the formation of glass becomes a true (singular) phase transition in the limit $T\to 0$. Similarly, using the Nagel-Schreckenberg model to simulate traffic flow, we show that the emergence of jammed traffic acquires the signature of a sharp transition in the deterministic limit $\pp\to 1$, corresponding to overcautious driving. We identify a true dynamical critical point marking the onset of coexistence between free flowing and jammed traffic, and demonstrate its analogy to the kinetically constrained glass models. We find diverging correlations analogous to those at a critical point of thermodynamic phase transitions.
\end{abstract}


\maketitle

Dynamic arrest, the sudden slow-down of dynamical systems with increasing density or interaction potential, is a central phenomenon in complex systems across biology, geology, material science, transport and traffic. The dynamic arrest is important for material stability and memory, but it is rather detrimental in traffic or transport, where congestion freezes any motion. A much studied example of dynamical arrest is the glass transition, i.e. the sharp increase of the viscosity of glass forming liquids. Many more systems exhibit similar dynamic arrest phenomena that appear to be related. An example is that of traffic flow: similar to the arrest of atomic motion in a glass, cars slow down and eventually arrest at high density due to crowding. While the similarity between different arrest phenomena has been pointed out, it is not clear to which extent they really are related.

A possible unifying scheme is that of a dynamic phase transition, in which the dynamic slow down is considered to be analogous to an equilibrium phase transition with its singularities in thermodynamic quantities. The central question then concerns the universality of the dynamic arrest. While signatures of dynamic phase transitions have been obtained in a few examples, the connection between these systems remains unclear, but would provide an important step to establish evidence of the universality of the dynamic arrest. With this letter, we aim to address this issue by demonstrating the presence of a dynamical critical point in a simple model for traffic flow, and connecting it to the dynamic slow-down of glasses.
{ Our goal is to establish a direct analogy between dynamic traffic flow models and the dynamics of glasses to highlight universal aspects of dynamic arrest.}

Important insight into the dynamic arrest of glasses comes from kinetically constrained models (KCMs), a class of discrete models with stochastic dynamics that are used to describe the glassy behavior and increasing relaxation time scales in supercooled liquids~\cite{KCMreview}. As the defining ingredient, KCMs have a kinetic constraint that allows local activity only if a local condition is met. These models provide some evidence that indeed the dynamic slow-down is the manifestation of a dynamic critical point~\cite{biroli,glassT0transition} in the limit $T\to 0$.

A constraint analogous to those in KCMs exists in traffic flow: cars can accelerate only if the distance to the car in front is sufficiently large. A well-studied model that incorporates a number of basic dynamical properties of real traffic is the Nagel-Schreckenberg (NS) model~\cite{NS}, a lattice-gas-like model with discrete position, time, and velocity. It allows stochastic fluctuations in the velocities of the individual cars, controlled by the probability $\pp$ that reflects the drivers' individual freedom to adjust their speed. The NS model describes the formation of traffic jams, and there has been much discussion regarding the (non)existence of a sharp phase transition between free-flowing traffic and traffic with jams~\cite{NSphasetransitionpolemic,Boccara,Chen}. There are a number of other models that describe traffic flow more realistically, including some that are extensions of the NS model~\cite{NSderivative-threephase}. Here, we focus on the most basic version of the NS model to demonstrate that already this simple one-dimensional model exhibits a true dynamical critical point allowing us to draw a direct connection to the dynamics of glasses. The key in the correspondence is the kinetic constraint, both present in KCM's of glasses and in all models of traffic in which collisions are avoided.
{ The beauty of our result lies in the fact that the relatively simple NS model of traffic fshows a dynamic phase transition that is surprisingly analogous to that of glasses.}

We consider the deterministic limit $\pp \rightarrow 1$ where cars tend to decelerate, and show that the NS model~\cite{NS} exhibits a non-trivial dynamical phase transition from free flow to coexisting free flow and jammed traffic. By applying dynamic correlation functions and susceptibilities normally used for glasses~\cite{biroli}, we show that at the onset of dynamic phase coexistence, the maximum correlation length and time diverge, analogous to the dynamic transition in KCMs at $T \rightarrow 0$. These simple stochastic dynamical systems allow for a direct comparison of dynamic arrest in different dimensions. We identify observables in traffic that behave similarly and that relate to each other in the same way as in glasses. This signals some degree of universality of the dynamic arrest.

The NS model simulates traffic flow in discrete space and time. A fixed number of cars with average density $\rho$ per lattice site are positioned on a one-dimensional lattice with periodic boundary conditions. The cars have integer velocities $v_i$ between 0 and some maximum velocity $v_\mathrm{max}$. The dynamics is given by the following update rules applied in parallel to all cars: All cars $i$ with velocity $v_i < v_\mathrm{max}$ accelerate by 1. Next, the kinetic constraint is applied, so that, if the distance of car $i$ to the next car $d_i < v_i$, then car $i$ decelerates to $d_i$. With probability $\pp$, a car reduces its velocity by 1. This parameter is the only source of stochasticity in the system and controls the velocity fluctuations. Finally, all cars move along the road by $v_i$ lattice sites.
In our simulations, we have propagated $2^{14}=16384$ cars under periodic boundary conditions.  Near $p=1$, where averages converge slowly, a steady state was produced by propagating the system for $5\times 10^7$ time steps.  Averages were calculated over a further $5\times 10^8$ time steps.

\begin{figure}
\includegraphics[height=5.75cm]{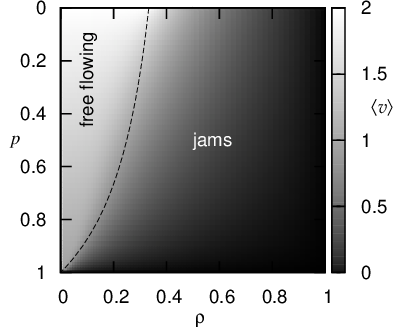}
\caption{
Phase diagram of the Nagel-Schreckenberg model showing $\langle v\rangle$ for the case of $v_\mathrm{max}=2$ (see gray scale on the right). The dashed line marks the transition between freely flowing traffic and traffic with jams shown in Eq.~(\ref{eq:transitionrhovmax2}).
}
\label{fig:phasediagram}
\end{figure}

We give an overview of the traffic flow in the NS model in Fig.~\ref{fig:phasediagram}, where we show the average velocity $\langle v \rangle$ as a function of $\rho$ and $\pp$ for $v_\mathrm{max}=2$. Traffic exhibits free flow at low density, where $\langle v \rangle = v_{\rm f} \equiv v_\mathrm{max} - \pp$, the velocity of free flow. With increasing density, cars interact and decelerate according to the kinetic constraint, leading to the formation of jams that coexist with free flow, and a concomitant decrease of $\langle v \rangle$. Here, a car is defined as being jammed if it has velocity zero. The transition density $\rho_{\rm tra}$ between free flow and traffic with jams can be estimated from the balance of the outflow and inflow rates of a jam, as required for its stability~\cite{gerwinksi}: The outflow rate of a continuous sequence of jammed cars is $1-p$, corresponding to the probability of acceleration of the car at the head of the jam. Because cars approach the rear of a jam with average velocity $v_{\rm f} = v_\mathrm{max}-p$, and the rear of the jam itself travels backwards at a speed equal to the inflow/outflow rate, this yields the transition density~\cite{gerwinksi}
\begin{align}
\rho_{\mathrm{tra}}
= \frac{1-p}{v_\mathrm{max}+1-2p}~,
\label{eq:transitionrhovmax2}
\end{align}
which reduces to $\rho_{\mathrm{tra}} \propto 1-p$ for small $1-p$. We use this relation to rescale the density near $p=1$.

How does the jammed regime emerge from that of free flow? For glasses and KCMs~\cite{KCMT0transition,KCMtransitions,glassT0transition}, it has been shown that the dynamic arrest becomes singular at $T=0$, where the dynamics become deterministic. The question is then whether a similar singular transition exists in the simple traffic flow model at $\rho \sim \rho_{\rm tra}$. To explore this analogy, we relate the stochasticity parameter $\pp$ to the temperature $T$ of spin glasses. The case $T = 0$, where the dynamics of glasses freezes entirely, corresponds to the case $\pp=1$, where cars always decelerate, and traffic flow arrests. In the limit of $T \rightarrow 0$ and $\pp \rightarrow 1$, the systems become deterministic.

\begin{figure}
\includegraphics[angle=270,width=8cm]{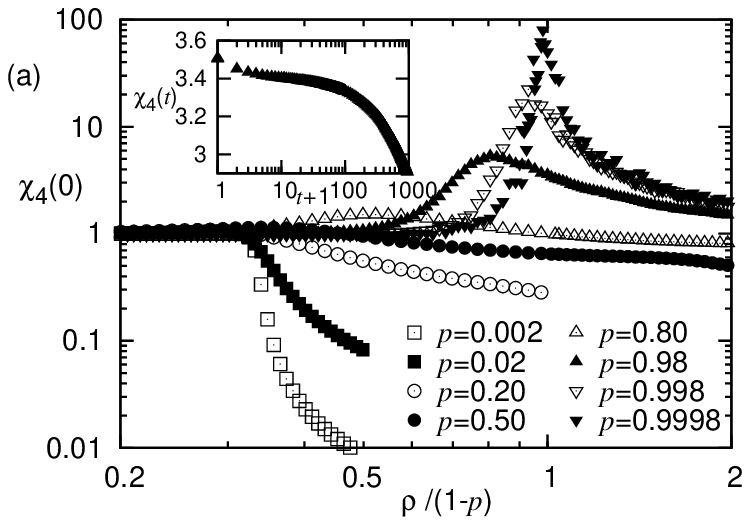}
\includegraphics[angle=270,width=8cm]{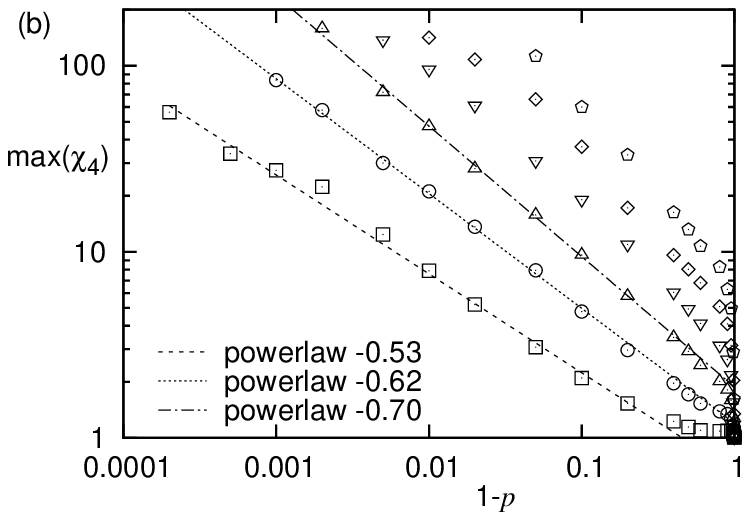}
\includegraphics[angle=270,width=8cm]{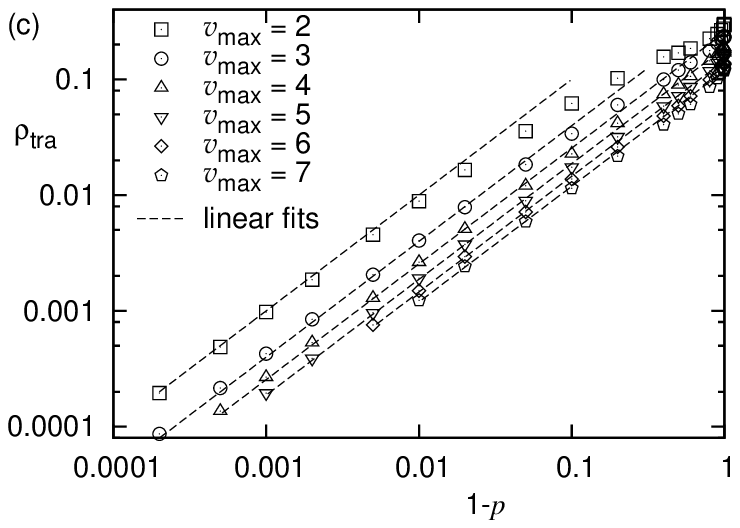}
\caption{
Dynamic susceptibility of traffic flow in the NS model. (a) The value of
$\chi_4(0)$ as a function of rescaled density for a range of
probabilities $\pp$ at $v_\mathrm{max}=2$. The peak sharpens markedly as
$\pp \rightarrow 1$. Inset: $\chi_4(t)$ as a function of time shows that
the largest value occurs at $t=0$.
(b) Maximum value of $\chi_4(t=0)$ as a function of the rescaled density,
plotted vs. $1-\pp$, for a set of
$v_{\rm max}$ (symbols explained in (c)). Power-law behavior (dashed lines) indicates the divergence of the dynamic susceptibility on approach of the critical point $\pp=1$. (c) Density of maximum dynamic susceptibility as a function of $1-\pp$ for various values of $v_\mathrm{max}$.
The position of the maximum is well described by the limiting behavior
of (\ref{eq:transitionrhovmax2}), $\rho_{\rm tra}=(1-\pp)/(v_{\rm
  max}-1)$ indicated by the dashed lines.
\label{fig:lengthscale}
}
\end{figure}

A characteristic property of the glass is its dynamic heterogeneity. Dynamically active regions separate from dynamically less active regions in space and time, leading to increasing dynamic heterogeneity of the system. This dynamic heterogeneity is quantified by the dynamic susceptibility~\cite{biroli,Glotzer1999_1,Glotzer1999_2}. In analogy, we define the dynamic correlation function of traffic flow using
\begin{align}
G_4(i,t) = \langle c(i;t) c(0;t) \rangle - \langle c(0;t)\rangle^2~,
\end{align}
where we take the mobility of car $i$ as $c(i;t) = (1/(t+1)) \sum_{t'=0}^{t} v_i(t')$, its average velocity during the time interval $[0,t]$. We find that near $\pp=1$, dynamic correlations become indeed increasingly long-ranged when the density approaches $\rho_{\mathrm{tra}}$. To investigate this increase in the correlation length, we define the dynamic susceptibility
\begin{align}
\chi_4(t) = \frac{1}{\langle v^2\rangle - \langle v\rangle^2} \sum_{i=0}^{N-1} G_4(i;t)~
\label{eq:chi4}
\end{align}
that measures the number of cars that move cooperatively on the time scale $t$. The dynamic susceptibility $\chi_4$ indicates the size of regions of correlated mobility, and has been much used to measure dynamic heterogeneity in glasses and granular materials
\cite{DynamicSusceptibility1,DynamicSusceptibility2,DynamicSusceptibility3,DynamicSusceptibility4}.
While in glasses, maximum cooperative motion arises at intermediate time scales, at which the particles escape their cages, in traffic, cars cannot escape the crowding of their environment independently from each other, and the maximum dynamic susceptibility arises at the shortest time interval, see inset of Fig.~\ref{fig:lengthscale}a.
To explore the growth of correlations, we focus on $t=0$, and show $\chi_4(t=0)$ as a function of density in Fig.~\ref{fig:lengthscale}a. Indeed, increasing maxima develop at $\rho \sim \rho_{\rm tra}$ as $\pp$ approaches unity, indicating increasing dynamic correlations. The divergence of the dynamic susceptibility is clearly seen in Fig.~\ref{fig:lengthscale}b, where we plot the maximum value of $\chi_4$ as a function of $1-\pp$. The figure shows data for various $v_{\rm max}$; in all cases, the maximum of $\chi_4$, $\chi_{4,{\rm max}} \propto (1-\pp)^{-\nu}$, indicating that the number of cars that move cooperatively diverges. In the NS model, the exponent appears to increase weakly with $v_{\rm max}$ changing from $\nu=0.53$ to $\nu>0.70$. This divergence indicates that traffic flow becomes truly critical in the limit $\pp \rightarrow 1$. The divergence is analogous to the one observed in KCMs at $T \rightarrow 0$, and indicates that the deterministic limits $\pp \rightarrow 1$ and $T \rightarrow 0$, are dynamical critical points of the systems. The divergence occurs at the onset of the jammed regime; to show this, in Fig.~\ref{fig:lengthscale}c, we compare the location of the maximum of $\chi_4$ (symbols) with the limiting ($\pp\to 1$) behavior of $\rho_{\rm tra}$ according to Eq.~\ref{eq:transitionrhovmax2} (dashes lines).

\begin{figure}
\includegraphics[angle=270,width=8.6cm]{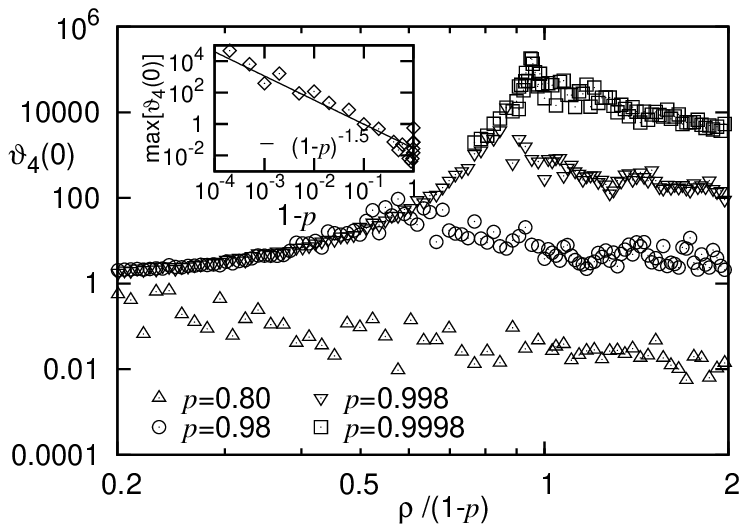}
\caption{
Residence time of cars in jams as a function of reduced density for various values of $\pp$ at $v_\mathrm{max}=2$. As $\pp \rightarrow 1$, the correlation time becomes more sharply peaked. Inset: Maximum residence time as a function of $1-\pp$ shows power-law divergence of the correlation time for $\pp \rightarrow 1$.
\label{fig:timescale}
}
\end{figure}

Further evidence of critical behavior comes from measurement of the correlation \textit{time} scale. To estimate the typical persistence time scale, we make use of a quantity similar to $\chi_4$, where we interchange time and car index in the definition of $c(i;t)$ and in the sum appearing in (\ref{eq:chi4}), to obtain $\vartheta_4(i)$. The temporal susceptibility $\vartheta_4$ indicates the correlation time scale of the system, and measures the typical residence time of a car in a jam. We plot this correlation time as a function of reduced density in Fig.~\ref{fig:timescale}. A strong increase of the maximum of $\vartheta_4$ suggests that in addition to the divergence of the correlation length, there is also a divergence of the correlation \textit{time} scale.
This is confirmed by plotting the maximum values of $\vartheta_4$ as a function of $(1-\pp)$ in the inset. Similar to the spatial correlations, the correlation time scale diverges as a power law $\vartheta_{4,{\rm max}}=(1-\pp)^{-\mu}$ as $\pp \rightarrow 1$, confirming that the system behaves critically along the time dimension. We determine the exponent to be $\mu \sim 1.5$. For real traffic, such diverging correlation time can have unpleasant consequences, as it indicates diverging persistence times of traffic jams.

We thus find a dynamical critical point characterized by diverging length and time scales. This critical point separates free flowing traffic from coexisting free flowing traffic and jams. This situation appears similar to equilibrium phase transitions, where the coexistence of phases is terminated by a critical point. To explore this analogy, we monitor the length of jams as a function of time, and find that indeed in the limit $\pp \to 1$, jams always coalesce in time to form a single jammed phase, coexisting with a single free flowing phase, analogous to the coarsening of equilibrium phases.

\begin{figure}
\includegraphics[angle=270,width=8.6cm]{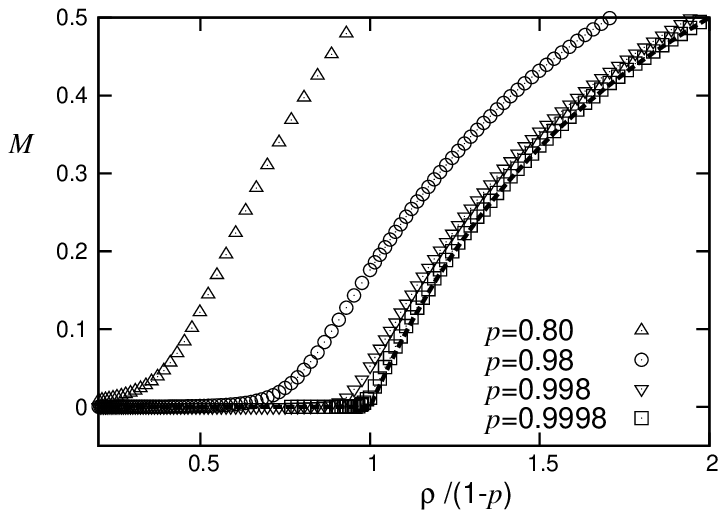}
\caption{
Order parameter $M$ as a function of reduced density for various values of $\pp$ at $v_\mathrm{max}=2$. As $\pp \rightarrow 1$, the transition at $\rho=1-\pp$ becomes singular. The dashed line shows the theoretical prediction $M=(\rho-\rho_{\rm tra}) / \rho$.
\label{fig:M}
}
\end{figure}

We explore this analogy further by defining the dynamical order parameter
\begin{align}
M = \frac{v_{\rm f} - \langle v \rangle}{v_{\rm f}}~,
\end{align}
the normalized deviation of the average velocity from that of free flow. We show $M$ as a function of the rescaled density in Fig.~\ref{fig:M}. It exhibits an increasingly sharp kink as $\pp \rightarrow 1$, but remains continuous at the transition $\rho = \rho_{\rm tra}$, indicating a singular point in the limit $\pp \to 1$.  If we assume simple coexistence in the two-phase regime, we can predict the function $M(\rho)=(\rho-\rho_{\rm tra})/\rho$, which we indicate as a dashed line in Fig.~\ref{fig:M}.  Indeed as $\pp\to 1$, there is strong evidence that the data  converge to this simple function, supporting our picture of jam and free flowing traffic as coexisting phases. The functional dependence of this order parameter has an exponent $\beta = 1$, corresponding to a Bose condensate, and to condensates found in typical zero range models \cite{zero-range}.

The dynamical phase transition we have described is analogous to that of directed percolation, as also indicated by \cite{KCMT0transition}, describing the transition at which an active (here: jammed) phase begins to permeate the entire system. We are in this stage, however, not prepared to say that the phase transition is in the universality class of directed percolation, in the sense that the critical exponents are the same. The apparently non-universal behavior of the exponent $\nu$ may indicate a more complex situation.

We have shown that the simple one-dimensional Nagel-Schreckenberg model for traffic flow already exhibits hallmarks of a dynamic phase transition analogous to that of kinetically constrained models for glasses. Exploiting this analogy, we have identified a dynamical critical point in the deterministic limit $\pp\to 1$ and $\rho\to 0$, marking the onset of coexisting jammed and free flowing traffic. The hallmark of this transition is the divergence of both correlation length and time scales, giving it a second order character. By defining a proper dynamic order parameter, we have shown that this transition is analogous to equilibrium phase transitions in which the coexistence regime is entered through the critical point. The direct analogy to KCMs of glasses points out the universality of dynamic arrest phenomena in systems of different dimensionality. Our results thereby offer a general scheme to comprehend dynamic arrest phenomena to be tested in other systems as well.

\end{document}